# Controlling thermal emission with metasurfaces and its applications

Qiongqiong Chu, Fan Zhong, Xiaohe Shang, Ye Zhang, Shining Zhu and Hui Liu [*]

National Laboratory of Solid State Microstructures, School of Physics, Collaborative Innovation Center of Advanced Microstructures, Nanjing University, Nanjing, Jiangsu 210093, China,

**Qiongqiong Chu and Fan Zhong** are contributed equally to this work.
*corresponding author: Hui Liu, E-mail: liuhui@nju.edu.cn.

**Abstracts:** Thermal emission caused by the thermal motion of the charged particles is commonly broadband, un-polarized, and incoherent, like a melting pot of electromagnetic waves, which makes it unsuitable for infrared applications in many cases requiring specific thermal emission properties. Metasurfaces, characterized by two-dimensional subwavelength artificial nanostructures, have been extensively investigated for their flexibility in tuning optical properties, which provide an ideal platform for shaping thermal emission. Recently, remarkable progress was achieved not only in tuning thermal emission in multiple degrees of freedom, such as wavelength, polarization, radiation angle, coherence, and so on but also in applications of compact and integrated optical devices. Here, we review the recent advances in the regulation of thermal emission through metasurfaces and corresponding infrared applications, such as infrared sensing, radiative cooling, and thermophotovoltaic devices.

## 1. Introduction

Thermal emission, a fundamental physical process of emitting electromagnetic energy spontaneously from objects with a temperature above absolute zero due to the thermal motion of charged particles, is experienced in daily lives ranging from the sunlight to burning candles. In the 19th century, Planck's law and Stefan-Boltzmann's law successively developed to theoretically describe the blackbody emission. Fundamentally, the thermal emission of a hot object is related to its emissivity $\varepsilon$ and temperature $T$ while the $\varepsilon$ of a blackbody is fixed as 1. Conventional thermal emission is broadband, un-polarized, and incoherent. Manipulating the temperature or the emissivity of the materials is the effective way to obtain desired emission spectrum. Therefore, precise control of a material's emissivity is essential for various thermal management applications. In the past 20 years, with the rapid development of nanophotonics and nanofabrication technology, the exploration of nanostructures at wavelength or sub-wavelength scales has brought the breakthrough for tuning thermal emission.

Various photonic structures, such as metasurfaces [1, 2], metamaterials [3, 4], photonic crystals [5-7], and multi-layer structures [8] have been widely explored to tune thermal emission. Compared with the latter three, metasurfaces characterized by two-dimensional subwavelength nanostructures with ultrathin thickness, can provide spatially tunable emissivity and are more suitable for the integration and miniaturization of infrared devices. In recent years, metasurface technologies have shown excellent capabilities in manipulating optical waves with multiple degrees of freedom including amplitude [9-11], phase [12-16], and polarization [17-19]. Many intriguing optical functionalities and applications, such as meta-lens [20, 21], biosensors [22, 23], and optical absorbers [24-28] have been achieved by metasurfaces. Artificial-designed metasurfaces provide a promising platform for tuning thermal emission. According to Kirchhoff's law, the emissivity of reciprocal thermal emitters is equal to their absorptivity under the thermal equilibrium condition. Consequently, controlling absorptivity is the key point for tuning thermal emission. Metasurface absorbers [29-32] have been widely investigated for one decade and are suitable candidates for the ideal thermal emitter. Among them, many metasurfaces such as grating [33, 34], micro-cavity [35, 36], periodically arranged metallic [37, 38], and dielectric [39, 40] meta-atoms have been proposed to manipulate the optical absorption and thermal emission.

Through reasonably designing the unit cell of metasurfaces, flexible tuning of thermal emission with multiple degrees of freedom, including emission wavelength, bandwidth, polarization, radiation angle, and spatial coherence has been demonstrated. To meet the requirement of developing infrared applications, various tuning mechanisms have been proposed, such as dynamic tunable thermal emitters incorporating active materials, nonreciprocal thermal emission that violates Kirchhoff's Law, pixelated thermal emitter array, and so on.

In this paper, we review the recent advances in thermal emission tuning based on metasurfaces, ranging from single metasurface to pixelated metasurface array. The multiple tuning degrees of freedom of thermal emission, including emission spectrum, polarization, radiation angle, and coherence are separately discussed in Sections 2-5. In Section 6, we discuss the research on dynamic tunable thermal emission. Then, the nonreciprocal thermal emission that violates Kirchhoff's law and near-field thermal emission are discussed in Sections 7 and 8. In Section 9, various infrared applications facilitated by thermal emission tuning are introduced. Also, the existing challenges for developing infrared devices are discussed. In Section 10, the research on pixelated thermal emitter array is discussed. The conclusions and future perspectives for thermal emission tuning on metasurfaces are discussed in Section 11.

## 2. Wavelength-selective thermal emission

Here, we first introduce the research progress of spectral thermal emission control. Blackbody emission, such as the emission from incandescent sources usually possesses a broadband emission spectrum covering the whole infrared wavelength range. Most emission energy goes into the unwanted infrared range and consequently causes low emission efficiency. Therefore, realizing rational wavelength-selective emission while suppressing the emission at other wavelengths as much as possible is essential for practical infrared applications, such as thermophotovoltaic and thermal management devices.

Various metasurfaces [43-44] composed of resonant meta-atoms have been proposed to achieve narrowband thermal emission. Among them, metal–insulator–metal (MIM) metasurfaces [43-46], composed of metallic meta-atom, metal mirror and insulator spacer between them, are frequently utilized to realize near-unity absorption through strong magnetic resonances. Thus, MIM metasurfaces are suitable for realizing the demanded thermal emitter. Liu et al. [47] demonstrated a MIM selective thermal emitter to realize narrowband infrared thermal emission, as shown in Figure 1(a). This thermal emitter consists of cross-shaped gold resonators on the top, a gold mirror on the bottom, and a silicon layer between them. Multi-resonant thermal emission can be further achieved by integrating multiple plasmonic resonators into one unit cell. Compared to narrowband resonance from MIM metasurfaces, even narrower emission peak is highly demanded for infrared sensing devices. By combining the intersubband transitions (ISB-T) in multiple quantum wells (MQWs) and photonic modes, De Zoysa et al. [48] experimentally achieved a very narrowband thermal emission, as shown in Figure 1(b). Recycled energy leads to enhanced emission intensity within the narrowband.

Broadband wavelength-selective thermal emission is another type of emission spectrum tuning, which is usually required by thermal management devices, for example, radiative cooling, thermal camouflage and thermal imaging. Through tapered metasurfaces [49-52] or by incorporating multi-resonance meta-atoms in metasurfaces [53-59], broadband thermal emission can be achieved. Argyropoulos et al. [49] demonstrated a broadband thermal emitter through tapered plasmonic Brewster metasurfaces, as shown in Figure 1(c). This broadband property comes from nonresonant Brewster funneling around Brewster's angle and the adiabatic focusing effect. Away from the Brewster angle, the emissivity of the metasurface shows a relative decrease. Further extending from 1D to 2D grating, this design is able to possess emission on all planes of TM polarization. In contrast to nonresonant design, multi-resonant metasurfaces possessing multiple optical modes typically enable maximum emission intensity in the normal direction. Zou et al. [53] demonstrated a low-cost metal-loaded dielectric metasurface to realize broadband thermal emission covering 8-13μm, as shown in Figure 1(d). The metasurface consists of a thick doped silicon substrate, two identical rectangular dielectric resonators, and a thin silver layer of 100 nm on the top surface. Two magnetic resonances at 8.8 and 11.3μm in dielectric resonators are the reasons for this wide-angle broadband emission spectrum.

Those works provide many metasurface technologies for tuning thermal emission spectrum. Proposed wavelength-selective thermal emitters can perform as efficient and low-cost infrared radiation sources, providing a new platform for various infrared applications.

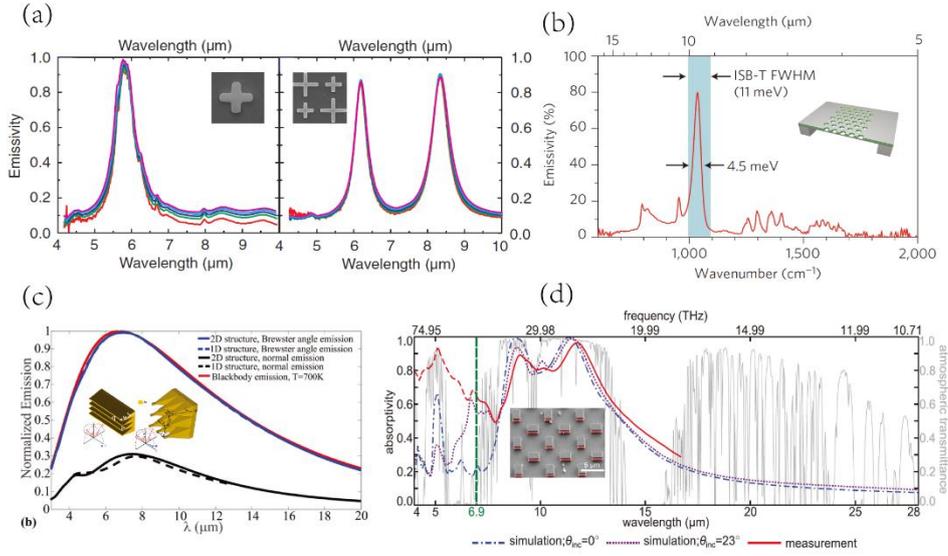

Figure 1. Wavelength-selective thermal emission. (a) Emission spectra of Au/Si/Au metasurfaces consisting of one or two resonators in a unit cell. Reprinted with permission [47]. Copyright 2011, American Physical Society. (b) Normally measured emission spectrum of MQWs with photonics designs under the temperature of 150℃. Reprinted with permission [48]. Copyright 2012, Springer Nature. (c) Comparison of the normalized emission spectra of 1D and 2D structures at T = 700 K. Reprinted with permission [49]. Copyright 2013, American Physical Society. (d) Measured broadband absorption/emission spectrum of the metal-loaded metasurface. Insets: the illustration of metasurface designs or the scanning electron microscope (SEM) images of fabricated metasurfaces. Reprinted with permission [53]. Copyright 2017, Wiley-VCH.

## 3. Angular dependence of thermal emission

To improve the emission efficiency, it is necessary to further restrict wavelength-selective thermal emission within demanded emission directions, which is important for many energy applications. Various metasurface-based mechanisms have been proposed to control the angular response of thermal emission, such as critical coupling [60], plasmon-phonon coupling [61] and Fano resonance [62].

Costantini et al. [60] proposed an Au/SiN/Au sandwich-like metasurface to realize narrowband and directional thermal emission, as shown in Figure 2(a). To improve the high-temperature stability of this MIM design, Au is changed to tungsten (W) in the fabricated sample. Measured emission spectrum shows that high-Q emission with half maximum (FWHM) of 200 nm is emitted within a limited solid angle (0.84 sr). An emission efficiency of 3.1% is realized. This MIM metasurface supports the gap surface plasmon mode, which satisfies the critical coupling condition in the normal direction, resulting in high emissivity at 4.25μm. When the incident angle increases, a new diffracted channel appears and then the critical coupling condition no longer be fulfilled, resulting in decreased emissivity. This is the reason for directional selective thermal emission.

By introducing mode coupling mechanisms [63-66], the tuning technologies of thermal emission can be further enriched. Through the coupling between angle-dependent and angle-independent modes, the angular response of thermal emission can be flexibly tuned. For example, the coupling between MIM resonance and phonon modes in polar dielectric can be utilized to tune the angular emission pattern. Zhang et al. [61] demonstrated a coupling mechanism within Al/SiO$_2$/Al metasurface to manipulate the bandwidth, polarization and angular properties of thermal emission, as shown in Figure 2(b). Under TE (E field parallel to grating) incidence, only phonon resonant peak exists, showing an angle-dependent emission pattern. In contrast, under TM (E field perpendicular to grating) incidence, magnetic resonance is excited and couples to the phonon resonance inside SiO$_2$ layer. Through plasmon-phonon coupling, the thermal emission of phonon is enhanced and shows a polarized, wide-angle emission pattern. We know that the coupling between a bright mode with strong far-field radiation and a dark mode with weak far-field radiation can result in a high-Q Fano resonance. This coupling method can be utilized in thermal emitters to facilitate narrowband thermal emission with regulated far-field emission patterns. Through Al/SiN/Al metasurface, Zhang et al. [62] proposed a Fano-type thermal emitter, as shown in Figure 2 (c). This Fano resonance is caused by the coupling between the parity-symmetric dark magnetic resonance mode and the angle-dependent surface lattice (SLR) mode. Through varying the structural parameters, the dark mode can be manipulated to approach SLR mode. In order to experimentally demonstrate the emission properties, the authors proposed an angle-resolved thermal emission spectroscopy (ARTES) technology through FTIR. The sample is fixed on the rotational heating stage with a minimum step of rotation being 0.2°. A polarizer is placed in front of the window to select desired polarized thermal emission. The emission signal from the rotating sample is collected through the heater window and then analyzed by FTIR. The angle-resolved thermal emission results are shown in Figure 2 (c), which clearly shows the occurrence of Fano resonance as the grating width increases, exhibiting narrowband and directional thermal emission properties.

Above mentioned ARTES technology, which is related to the angular response of thermal emitter, is a useful tool for spectroscopy characterization. By utilizing the ARTES technology, Zhong et al. [67] directly characterized the dispersion of designed metacrystals, demonstrating intrinsic eigenmode properties of non-Hermitian systems, as shown in Figure 2(d). The top figure shows the ARTES measurement equipment. The bottom shows the designed metacrystal and measured thermal emission results at $h_b$ = 0.8755μm, which demonstrates the existence of EPs (Exceptional points). Also, this ARTES technology can be utilized to probe mid-infrared surface wave emission. Zhong et al. [68] demonstrated the existence of surface wave state on designed planar metacrystal through measured thermal emission dispersion, as shown in Figure 2(e). This ARTES technology can be applied to other spectroscopy analysis devices.

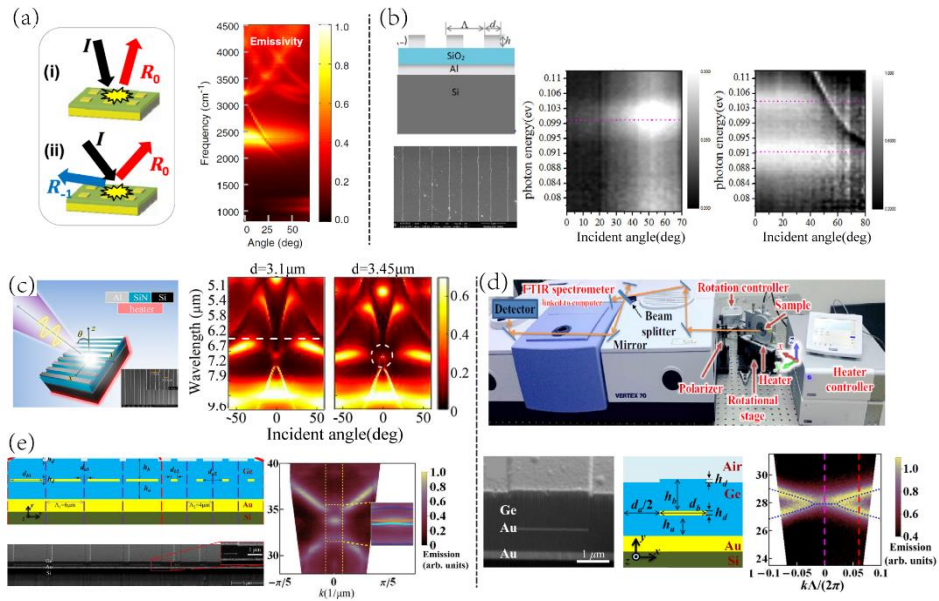

Figure 2. Angular properties tuning of metasurfaces. (a) The illustration of the absorption principle of designed Au/SiN/Au metasurface. And measured angle-dependent thermal emission of W/SiN/Pt metasurface. Reprinted with permission [60]. Copyright 2015, American Physical Society. (b) Designed Au/SiO$_2$/Au metasurface. And measured angle-dependent thermal emission under TE and TM polarization respectively. Reprinted with permission [61]. Copyright 2017, The Author(s). (c) Sketch plot of Au/SiN/Au metasurface. Inset: the SEM picture of the fabricated sample. And measured angle-resolved thermal emission with increased grating width d. The white line and the circle indicate the dark MR mode and the fano resonance point. Reprinted with permission [62]. Copyright 2019, American Chemical Society. (d) The top: ARTES measurement setup. The bottom from left to right: the SEM picture of the fabricated sample; the schematic of designed meta-crystals; Measured and calculated thermal emission dispersion. Reprinted with permission [67]. Copyright 2020, American Physical Society. (e) The left: the sketch and SEM picture of the designed superlattice. The right: measured and calculated thermal emission dispersion, indicating the existence of interface states. Reprinted with permission [68]. Copyright 2021, © The Optical Society.

## 4. Polarized thermal emission

### 4.1 Linearly polarized thermal emission

In contrast to un-polarized blackbody thermal emission, metasurface-based thermal emitters with fabricated subwavelength meta-atoms typically emit polarized thermal emission. For instance, antenna designs can generate linearly polarized thermal emission parallel or perpendicular to the antenna. Schuller et al. [69] experimentally demonstrated a SiC antenna thermal emitter, which can perform as a narrowband infrared radiation source, as shown in Figure 3(a). Multiple dielectric-type and metallic-type resonances of the designed antenna can be categorized as TE and TM polarized modes, exhibiting linearly polarized thermal emission. The polarization and resonant wavelength characteristics of antenna emitters can be well defined through size control. By manipulating the dissipative loss or bending the antenna resonators, the emission power of antenna emitters can be maximized toward the blackbody emission limit [70]. Except for antenna emitters, many metasurfaces such as grating [71, 72] and rectangular patches [44], can also be utilized to realize the linearly polarized thermal emission due to their resonance-induced high absorption under linearly polarized incidence. Mason et al. [72] proposed a wavelength-selective thermal emitter composed of gold grating, a dielectric layer and gold mirror, as shown in Figure 3(b). The antiparallel surface currents in the grating and the metal mirror cause the TM polarized thermal emission.

### 4.2 Circularly polarized thermal emission

In addition, circularly polarized thermal emission is another hot topic for polarization tuning of thermal emission. Since the chiral response is very weak in natural materials, metasurfaces [73-76] including chiral and achiral types, provide an effective way to realize circularly polarized thermal emission through diverse mechanisms such as Weyl semimetals with intrinsic nonreciprocal, hyperbolic materials, chiroptical effect from chiral morphology nanostructure, Rashba effect and so on.

Dahan et al. [76] experimentally displayed circularly polarized thermal emission using an achiral metasurface. A spin-split dispersion is obtained through a rotating antenna array. Corresponding spin-dependent emission spectra are shown in Figure 3(c). Chiral metasurfaces [77, 78] are another common way to efficiently enhance the chiral response by introducing symmetry breaking. Circularly polarized thermal emission can be emitted from those chiral metasurface absorbers and emitters. Utilizing a chiral silicon metasurface, Wu et al. [79] achieved a high-Q chiral absorption/emission through Fano resonance in dielectric resonators, as shown in Figure 3(d). This chiral metasurface possessing high degree of circular polarization (DCP) is a promising candidate for selective circularly polarized thermal emitter. For practical infrared spectroscopy applications, an electrically tunable compact infrared radiation source is highly demanded. Nguyen et al. [80] demonstrated an ultrathin incandescent chiral metasurface composed of connected Z-shaped resonators to realize circularly polarized broadband thermal emission, as shown in Figure 3(e). The emission signal from the chiral metasurface can be modulated beyond 10 MHz with a DOP (degree of polarization) larger than 0.5 from 5 to 7 μm. This design is suitable for compact incandescent radiation sources. Up to now, most proposed circularly polarized thermal emission is generated in the normal direction or exhibits a directional emission pattern. To overcome this obstacle, Wang et al. [81] demonstrated a symmetry-broken chiral metasurface to realize omnidirectional circularly polarized thermal emission with nonvanishing optical helicity, as shown in Figure 3(f). Through simultaneously breaking the mirror symmetry and inversion symmetry, asymmetric circularly polarized thermal emission can be realized. This symmetry-broken technology extends the tuning degrees of freedom of thermal emission.

Those chiral and achiral metasurfaces with ultrathin thickness provide the guideline for designing on-chip thermal emitters with linearly or circularly polarized thermal emission. The tuning ability of the polarization of thermal emission is important for infrared radiation sources and thermal sensing applications.

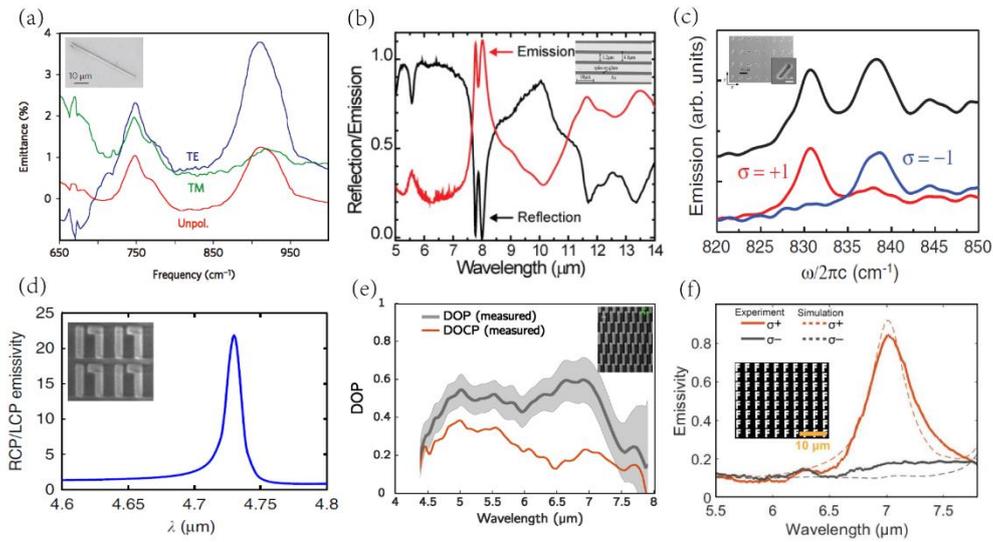

Figure 3. Polarization control of metasurfaces. (a) The measured emission spectra of antenna emitter under TM, TE polarization and un-polarized condition. Reprinted with permission [69]. Copyright 2009, Springer Nature. (b) Measured emission and reflection spectra of designed grating under TM polarization. Reprinted with permission [72]. Copyright 2011, AIP Publishing. (c) Measured emission spectra of thermal antenna array without a polarizer (black line), with an RCP (blue line) and LCP circular polarizer (red line). Reprinted with permission [76]. Copyright 2019, American Physical Society. (d) Obtained DCP of thermal emission of designed dielectric chiral metasurface. Reprinted with permission [79]. Copyright 2014, Springer Nature. (e) Obtained DOP and DOCP of incandescent chiral metasurface from measured circularly polarized thermal emission. Reprinted with permission [80]. Copyright 2023, © The Optical Society. (f) Measured emission spectra of designed F-shape chiral metasurface under LCP and RCP polarization. Insets: the SEM images of fabricated metasurfaces. Reprinted with permission [81]. Copyright 2023, © The Authors, some rights reserved; exclusive licensee AAAS. Distributed under a CC BY-NC 4.0 license http://creativecommons.org/licenses/by-nc/4.0/.

## 5. Coherent thermal emission

The researches above have shown the thermal emission tuning on emission spectrum and polarization characteristics. In addition to these two emission properties, the coherence characteristic is also an important aspect of tuning thermal emission. We know that, the emission from a thermal source is usually incoherent and its intensity is the sum of the emission intensities from different points, causing a non-directional emission pattern. Coherent thermal emission is the key point for realizing directional thermal emission, thermal focusing and holography. The former has been experimentally achieved while the latter two are still challenging in experiments. It should be noted that the correlated phase of local components of the metasurface is discussed in the research on directional thermal emission in this section, which is ignored in Section 3. The spatial coherence can be effectively enhanced through coupling to non-local optical modes, such as surface-phonon polaritons (SPhPs) [82-84], surface-plasmon polaritons (SPPs) [85, 86] or their hybrid modes while the temporal coherence [87, 88] is usually related to the emission bandwidth.

In 2002, Greffet et al. [89] pioneeringly demonstrated a coherent thermal source through SiC grating, as shown in Figure 4(a). Thermally excited near-field SPhPs enhance the coherence of emitted thermal light and thus enable directional thermal emission. Thereafter, abundant research on coherent thermal emission ranging from directional thermal emission to theoretically demonstrated thermal focusing has been proposed based on various metasurface technologies including grating [90-92], metallic grooves [93], and periodical meta-atom array [94]. Gratings made of polar material SiC enable SPhP modes within the Reststrahlen band, which have been extensively investigated. Normally, grating emits thermal light in an angular range (Δθ) due to angular dispersion. Ref [95] demonstrated that the angular dispersion originate in material dispersion and the misalignment between the dispersion curve of the grating and the light line, and then provided a solution to this problem. A SiC grating combined with bundled graphene sheets is proposed to alleviate the two aforementioned limitations and thus achieve a reduction of angular dispersion from 30° to 4° within the SPhPs wavelength range of 11–12 μm. Thermally excited SPPs are another common way to enhance the coherence without wavelength range limitation. Ref [96] theoretically demonstrated that periodically distributed circular metallic grooves are able to emit a narrowband beam in the normal direction within a narrow angular range. Subsequently, Park et al. [97] experimentally verified the feasibility of this theoretical design in realizing highly directional thermal emission. A series of circular concentric grooves are fabricated on W and molybdenum (Mo) films, and the fabricated W bull's eye is shown in Figure 4(b). Spectrally narrow and highly directional (~2°) thermal beam is emitted at 900 °C through coupling to thermal excited SPPs. Typically, SPPs in optical metasurfaces suffer from metal loss. However, this feature will turn into an advantage for thermal devices due to enhanced absorption and emission of metallic parts. Hybrid surface plasmon-phonon-polaritons (SPPhPs) [61, 98, 99] or the strong coupling between SPhPs/SPPs [100] and other optical modes can further enrich the tuning mechanisms of coherence properties of thermal emission.

Partial coherence can lead to directional thermal emission while the complete thermal emission control over amplitude, polarization, local phase, and coherence is still difficult. Overvig et al. [101] theoretically proposed a thermal metasurface technology to address this challenge, which requires both local and nonlocal modes, as shown in Figure 4(c). The local modes enable wave front shaping while the nonlocal modes add coherence to thermal emission. Specifically, ultrathin two-layer dielectric metasurfaces possessing symmetry-controlled nonlocal QBIC modes and local modes are theoretically designed to achieve thermal emission focusing and wave front control with designed spin and angular orbital momenta. Up to now, well-defined wave front control of thermal emission for realizing thermal emission focusing [102] and thermal holography [103] still lack experiments.

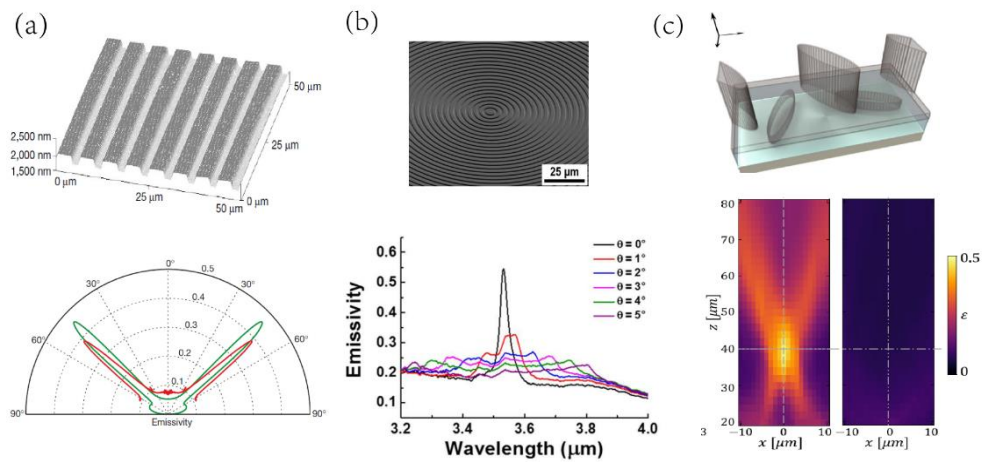

Figure 4. Coherent thermal emission. (a) The top: the image of the SiC grating obtained by atomic force microscopy. The bottom: polar plot of measured emissivity of the grating at λ = 11.36μm. Reprinted with permission [89]. Copyright 2002, Springer Nature. (b) The top: the SEM picture of fabricated W bull's eye sample. The bottom: measured emissivity of fabricated bull's eye under varied emission angle at 900 °C. Reprinted with permission [97]. Copyright 2016, American Chemical Society, https://pubs.acs.org/doi/10.1021/acsphotonics.6b00022. (c) The top: the schematic of Si-made thermal metasurface composed of structured silicon elliptical pillars monomers and dimers on the top and bottom layer respectively. The monomers work as the local elements while the dimers work as the non-local elements. The bottom: focusing of emitted thermal light at the x-z plane. Reprinted with permission [101].

## 6. Dynamic tunable thermal emission

Plenty of metasurface technologies have been explored to achieve precise control of thermal emission properties ranging from emission spectrum to coherence. However, most metasurface designs are static and their emission properties are fixed by the structural parameters. In recent years, dynamic tunable thermal emitters possessing switchable thermal emission properties under high-speed modulation have attracted extensive attention for the development of adaptive thermal management devices, such as spatially reconfigurable thermal imaging and adaptive thermal camouflage. Dynamic tuning of the emission properties is commonly realized by utilizing active materials including graphene[104-111], InAs [112], GaAs [113], GaN [114] and various phase change materials (PCMs), such as vanadium dioxide ($VO_2$) [115-119], GST ($Ge_2Sb_2Te_5$)[120-122] and so on[123, 124]. Correspondingly, many dynamic tuning mechanisms are explored including electrical modulation [125-127], temperature modulation [128, 129], ultraviolet (UV) irradiation [130], femtosecond laser pulse excitation and heating [131, 132] and so on.

Graphene, which possesses low thermal mass and high thermal conductivity, is a promising electrothermal material for on-chip and high-speed thermal emitters. Multilayer graphene can perform as an incandescent thermal source to provide a continuous infrared emission spectrum. Graphene resonators or metallic metasurface combined with graphene layers are capable of supporting plasmonic resonances in the infrared range for realizing wavelength-selective thermal emission. Brar et al. [105] experimentally demonstrated an electrically controllable narrowband thermal emitter composed of graphene antennas, $SiN_x$ membrane, and Au mirror. The top view of graphene antennas is shown in Figure 5(a). The emission wavelength and intensity of antennas are continuously tuned by varying the carrier density of graphene. Active InAs, GaAs, and GaN materials, as electrically tunable layers have also been utilized to realize dynamic thermal emission tuning. For example, Ref [114] demonstrated an electrically tunable narrowband mid-infrared thermal emitter through GaN/AlGaN multiple quantum wells with high modulation speed (50 kHz) under high temperature up to 500 °C.

GST and $VO_2$ are widely utilized PCMs for dynamic thermal engineering due to their ultra-fast modulation speed and repeatable cycles to temperature. Phase-change $VO_2$ material, which undergoes insulator-to-metal transition around 68 °C, is suitable for optoelectronic devices that require continuous modulation and cannot withstand high temperature. Chandra et al. [116] experimentally demonstrated an adaptive infrared camouflage system by incorporating $VO_2$ layer into cavity-LSPR (localized surface plasmon resonance) hybrid plasmonic metasurface, as shown in Figure 5(b). The thermally driven $VO_2$ layer enables the surface emissivity variations, thus facilitating multispectral and spatial infrared information encoding of thermal camouflage. In contrast to $VO_2$ material, GST material requires high temperature for crystalline-to-amorphous transition. However, once the phase transition is completed, the amorphous or crystalline state can be maintained for many years at room temperature, facilitating energy-efficient thermal devices. For example, Ref [120] proposed a zero-static-power mid-infrared thermal emitter to realize spectrally dynamic control of thermal emission by utilizing GST layers.

Through combining the active materials and reconfigurable tuning mechanisms, the tuning degrees of freedom of thermal emitters are greatly enriched, enabling simultaneously spectral, spatial, and temporal thermal emission control. Coppens et al. [130] experimentally demonstrated that spectral and spatiotemporal emissivity control can be realized by combining the MIM metasurface with a photosensitive ZnO layer, as shown in Figure 5(c). Under pixel-level spatial control of UV light, free carriers generated in the ZnO layer enable the temporally modulated emissivity of the metasurface. However, the emissivity modulation realized by this tuning method is limited to 0.12. To improve the modulation performance, Xu et al. [131] recently proposed a nonvolatile reconfigurable emissivity-coding metasurface, as shown in Figure 5(d). The mid-infrared thermal emission and visible scattering properties can be respectively modulated by laser-crystallized GST spots and laser-induced microscale bumps. This spatial combination of multispectral modes enables the anticounterfeiting functions. Demonstrated spectral and spatiotemporal thermal emission tuning with emissivity modulation up to 0.6 is reversible with zero static power consumption, facilitating nonvolatile thermal encryption.

Incorporating various active materials and external tuning mechanisms, such as laser heating and electrical modulation, enables dynamic thermal emission tuning with multiple degrees of freedom. Further enriching the tuning degrees and improving the modulation speed are the key points for the next generation of adaptive thermal management devices.

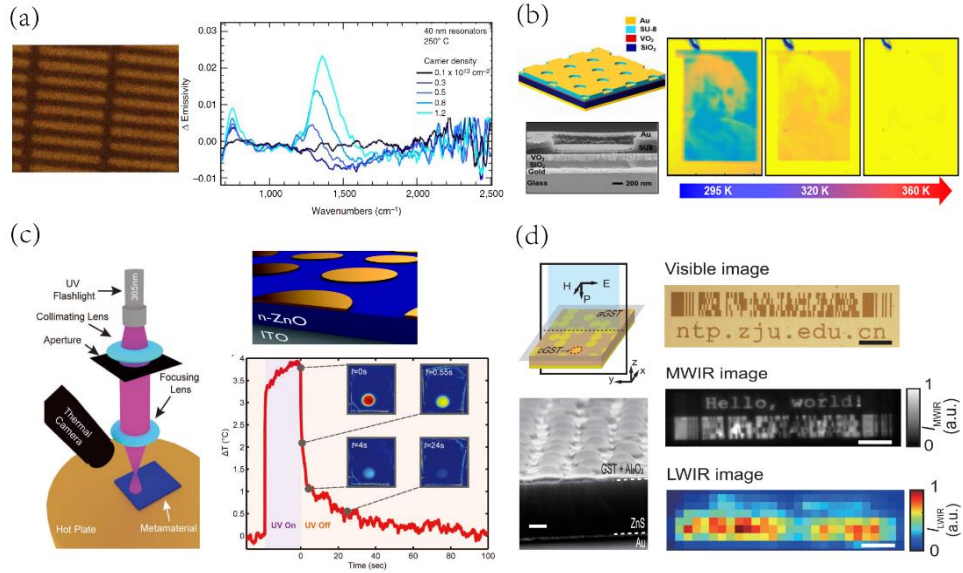

Figure 5. Dynamic tunable thermal emission. (a) The SEM image of fabricated graphene nanoresonators on a 1 mm thick SiNx membrane, and their emissivity variations with varied carrier density of graphene. Reprinted with permission [105]. Copyright 2015, Springer Nature. (b) Designed cavity-coupled plasmonic metasurface. And the tunable Albert Einstein image with increasing temperature. Reprinted with permission [116]. Copyright 2018, American Chemical Society. (c) The left: the measurement setup. The right: the sketch of the Au/Zno/ITO design and the temporal response of spatially illuminated spot recorded by the temperature variations. Reprinted with permission [130]. Copyright 2017, Wiley-VCH. (d) The left: the schematic shows the emissivity modulation by controlling the density of cGST points. The SEM image shows the submicron-sized bumps fabricated by ns laser pulses. The right: the binary anticounterfeiting label in different wavelength ranges. Reprinted with permission [131]. Copyright 2021, American Chemical Society.

## 7. Nonreciprocal thermal emission

Kirchhoff's law describes that the absorptivity and emissivity at a given wavelength ($\lambda$), polarization (p), and direction ($\theta$) are equal, $\alpha(\lambda, p, \theta) = e(\lambda, p, \theta)$. This law originates from Lorentz reciprocity of Maxwell's equations. Breaking the reciprocity can bring the nonreciprocal thermal emission/absorption that violates Kirchhoff's Law. Reciprocal absorbers suffer from the intrinsic loss coming from the unavoidable radiation back to the sun. Therefore, the nonreciprocal system [133-136] has fundamental importance for solar energy harvesting systems to reach their efficiency limit, known as the Landsberg limit. Also, nonreciprocal emission is appealing for thermal management devices, which are usually designed following Kirchhoff's law.

Under a strong magnetic field, MO (magneto-optical) materials [137, 138] that can provide asymmetric permittivity tensor, are suitable for achieving nonreciprocal thermal emission. In 2014, Zhu et al. [139] theoretically proposed a magneto-optical InAs grating to achieve a near-complete violation of Kirchhoff's law, as shown in Figure 6(a). At a strong magnetic field of 3 T along the z-axis, the emissivity and absorptivity under critical coupling conditions can reach a near-unity difference. To reduce the dependence of the nonreciprocal system on the magnetic field, Ref [140] demonstrated that guided mode in loss-less grating can efficiently enhance the MO effect, thus lowering the requirement on the magnetic field to 0.3T. Then, magnetic Weyl semimetals [133, 141, 142] due to their inherent time-reversal symmetry breaking and large anomalous Hall effect have become a hot spot for violating Kirchhoff's law. Zhao et al. [141] demonstrated a nonreciprocal thermal emitter composed of magnetic Weyl semimetal grating, as shown in Figure 6(b). Without requiring any external magnetic field, non-zero off-diagonal contributions of permittivity tensor can be obtained to realize nonreciprocal thermal emission. Since $e(\lambda, \theta) = \alpha(\lambda, -\theta)$, the nonreciprocal thermal properties are obviously indicated in the angular-dependent absorptivity. This breaking way of reciprocity is fundamentally different from that of MO materials. In addition to MO materials and magnetic Weyl semimetals, spatiotemporal modulation is another way to violate Kirchhoff's law through asymmetric scattering matrix. The spatiotemporal modulation of the antenna [143] and gratings [144, 145] have been proposed to achieve nonreciprocal effect in megahertz and long-wave infrared range. Ghanekar et al. [145] theoretically demonstrated nonreciprocal thermal emission based on the space−time modulation (modulation on temporal frequency Ω and spatial frequency G) of the Fermi level of graphene, as shown in Figure 6(c). In the presence of loss, asymmetric coupling leads to unequal absorption matrix A and emission matrix E, thus realizing the violation of Kirchhoff's law.

Above researches are investigated theoretically without experiments. Realizing the violation of Kirchhoff's law in experiments is difficult due to weak nonreciprocal effects caused by material loss in natural MO materials. Recently, Shayegan et al. [146] experimentally observed the violation of Kirchhoff's law from measured absorption and emission spectrum, as shown in Figure 6(d). Through coupling the guided-mode resonance (GMR) of dielectric grating to MO material InAs, the symmetry of permittivity tensor can be broken under magnetic field tuning. When the applied magnetic field is switched from +1.0 T to -1.0 T, measured emissivity and absorptivity changes satisfy the relationship of $\Delta e \cong -\Delta \alpha$, demonstrating the nonreciprocal emission properties for the same angle. At the ENZ (Epsilon-near-zero) wavelength of n-InAs, the nonreciprocal effect is strongest and can be extended to other wavelength ranges. Broadband non-reciprocal absorber/emitter under a moderate magnetic field is more appealing for thermal management devices. Liu et al. [147] experimentally demonstrated a broadband mid-infrared nonreciprocal absorber under a moderate magnetic field. The nonreciprocal effect comes from the excited Berreman modes (BMs) in InAs film and explored gradient ENZ properties. Moreover, various metasurface technologies can be incorporated to provide a tunable absorption spectrum for nonreciprocal absorbers. This design is expected to realize nonreciprocal thermal emitters.

These works lay the foundations for designing nonreciprocal systems for plenty of infrared applications, such as thermophotovoltaic devices, solar energy harvesting and thermal camouflage.

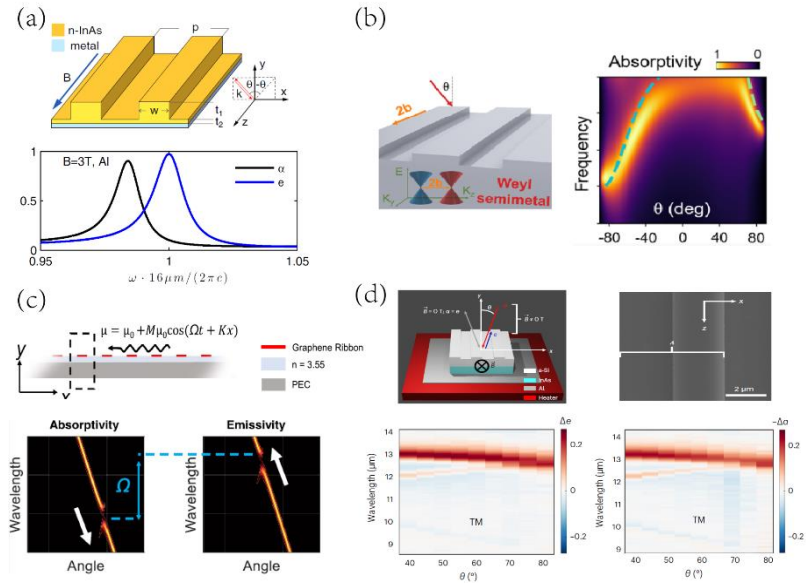

Figure 6. Nonreciprocal thermal emission. (a) The top: sketch of the nonreciprocal thermal emitter composed of InAs grating and a metal layer. The bottom: calculated absorptivity and emissivity spectra of the nonreciprocal thermal emitter at the condition of θ = 61.28 and B = 3 T. Reprinted with permission [139]. Copyright 2014, American Physical Society. (b) The left: schematic of the Weyl semimetal made grating. The right: the angle-dependent absorptivity, showing clear asymmetric properties. Reprinted with permission [141]. Copyright 2020, American Chemical Society. (c) The top: graphene grating based nonreciprocal design. The bottom: calculated absorptivity and emissivity under the spatiotemporal modulation of graphene Fermi energy. Reprinted with permission [145]. Copyright 2023, American Chemical Society. (d) The top: schematic of the GMR structure on the InAs layer. The bottom: the absolute change of measured emissivity and absorptivity as the magnetic field is switched from +1.0 to −1.0 T: $\Delta e = e^{-1.0\,T} - e^{1.0\,T}$ and $\Delta \alpha = \alpha^{-1.0\,T} - \alpha^{1.0\,T}$. Reprinted with permission [146]. Copyright 2023, The Author(s), under exclusive licence to Springer Nature Limited.

## 8. Near-field thermal emission

Thermal emission is one of the three fundamental channels in heat exchange, which propagates without a medium. Abundant works shown above have been done in studying far-field thermal emission, which is bounded by the Planck thermal-emission limit. However, subwavelength thermal emitters appear to exceed the limit, wherein some counterintuitive phenomena result from the difference between emission cross-sections and geometric cross-sections, and others occur when a temperature gradient is inside the internal energy distribution [148, 149].

Near-field thermal emission has been theoretically studied in many systems, including patterned 1D [150-156], 2D [156-158] nanostructures and 2D materials based designs, like graphene plane [159-161] and hyperbolic nanostructures [162-165]. The thermal fluctuations induce the evanescent waves inside the nanoscale gap to transfer the near-field radiative energy between nanostructures, where metasurfaces can provide solutions for the design of the thermal structure like gratings that have larger radiative flux than counterpart plane plates, exceeding blackbody limit. Liu et al. [157] theoretically investigated the near-field properties of patterned metasurfaces, as shown in Figure 7(a). The coupling between the hyperbolic modes supported by patterned 1D and 2D metasurfaces and the evanescent waves in the vacuum enables the enhancement of near-field heat transfer compared to unpatterned thin film. The near-field heat flux is specifically explored with varied metasurface parameters, including the volume filling ratio, period, and thickness. However, this near-filed heat transfer is smaller than that in parallel plates made of polar dielectrics. To address this problem, Fernández-Hurtado et al. [158] theoretically demonstrated a near-field radiative heat transfer method based on Si metasurfaces, which is much larger than its unpatterned counterpart and planar polar dielectrics at room temperature, as shown in Figure 7(b). Holes introduced in the Si layers lead to broadband SPPs, thus enabling the enhancement of near-field radiative heat transfer over a broad range of separations (from 13 nm to 2 μm).

Recently, graphene has been widely applied to near-field heat transfer system due to its highly tunable carrier mobility and strongly confined SPPs. For instance, "ON" and "OFF" switching states of heat conductance can be realized in the stacked graphene sheets [159, 160]. When constructing metasurfaces with 2D materials, the near-field heat transfer between nanoparticles placed above an array of graphene strips can be significantly enhanced [165]. Typically, the tunability of the hyperbolic phonon polaritons (HPPs) supported by hexagonal boron nitride (hBN) is limited. To address this problem, graphene-hBN hybrid designs [163, 164] are proposed to achieve tunable and enhanced near-field thermal emission. In addition to the intrinsic natural hyperbolic properties of hBN, multilayer graphene-hBN structure leads to more effective hyperbolicity, thus providing better performance in enhancing the near-field heat transfer [163]. Meanwhile, the varied chemical potentials of graphene enable the tunability of the design.

To observe the near-field thermal emission, readable near-field thermal emission signals are necessary in the experiment [166-170]. One of the solutions is to extract the near-field signals by a scattering-type scanning near-field optical microscope (s-SNOM). Ref [166] detected a sample of 300K through s-SNOM method without external illumination. A high spatial resolution of ~60 nm of near-field signals is achieved and the near-field images are enabled free from in-plane spatial interference. By guiding the near-field thermal emission signal emitted from sample to FTIR spectrometer, near-field thermal emission spectroscopy can be obtained [167]. Another approach based on a high-precision micro-electromechanical system (MEMS) offers an opportunity to quantitatively reveal the near-field radiative heat transfer between nanobeams. Through MEMS-controlled SiC nanobeams, St-Gelais et al [168]. demonstrated a near-field heat transfer enhancement two orders of magnitude larger than the far-field limit under the condition of ~42 nm separation between nanobeams, as shown in Figure 7(c). Under a separation smaller than 200 nm, the heat transfer follows the typical $1/d^\alpha$ law. High tensile stress on nanobeams enables the stable control of nanometer-scale separation even at large thermal gradients. Moreover, Yang et al [169]. proposed a homemade setup that can perform well in near-field detection, directly demonstrating the plasmon-mediated near-field thermal emission between two slices of macroscopic graphene predicted by theory, as shown in Figure 7(d). This measurement system can hold the smallest gap distance of around 430±25nm between two graphene sheets. Measured results of graphene sheets show 4.5 times larger heat flux density than the blackbody limit, exhibiting a clear Super-

Planckian radiation.

The exploration of blackbody limit in near-field thermal emission opens new avenues of applications for metasurfaces in energy harvesting and thermal management, such as high-resolution near-field imaging, thermal rectifiers, thermal transistors, and enhancing the conversion efficiency of thermophotovoltaic devices. Also, the near-field thermal emission is crucial for the thermal control issue in integrated nanoelectromechanical systems and thermotronics devices [170].

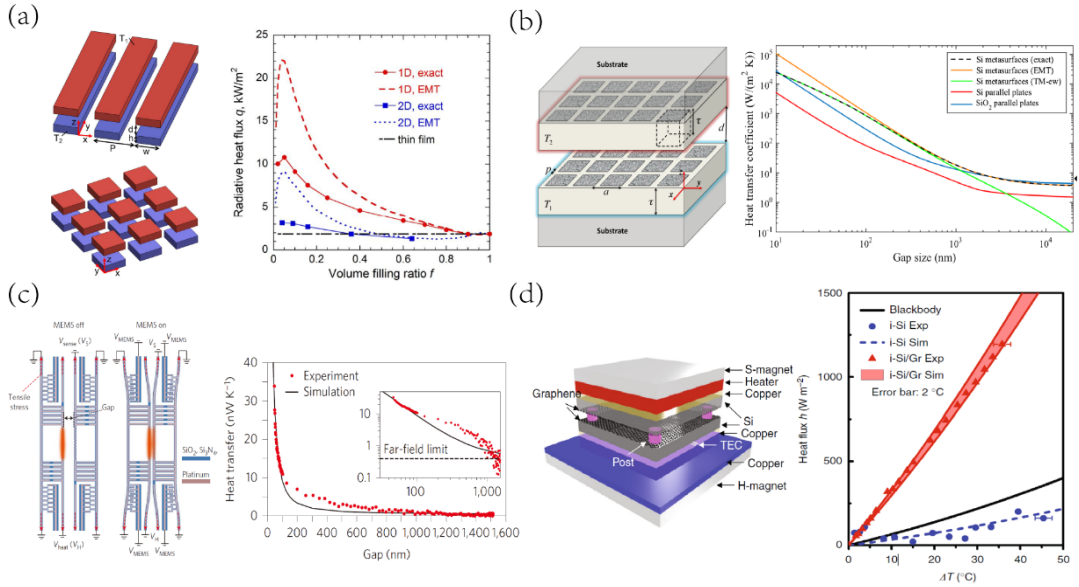

Figure 7. Near-field thermal emission. (a) The left: schematic of the near-field heat transfer of 1D and 2D metasurfaces. The right: the radiative heat flux of 1D, 2D metasurfaces and un-patterned thin film as a function of the volume filling ratio. Reprinted with permission [157]. Copyright 2015, American Chemical Society. (b) The left: schematic of two doped-Si metasurfaces for near-field heat transfer. The right: the room-temperature heat transfer coefficients of Si metasurfaces, Si and $SiO_2$ parallel plates as a function of the gap size. Reprinted with permission [158]. Copyright 2017, American Physical Society. (c) The left: schematic of the MEMS system. The right: measured and simulated heat transfer results between nanobeams as the separation gap increases. Inset: the same data on a logarithmic scale. Reprinted with permission [168]. Copyright 2016, Springer Nature. (d) The left: the schematic of the home-made measurement setup. The right: measured heat flux density (symbols) and the theoretical predictions (shaded region and dashed line) between i-Si substrates with or without graphene. Reprinted with permission [169]. Copyright 2018, The Author(s).

## 9. Infrared applications based on thermal emitters

Numerous researches on controlling thermal emission have brought new perspectives for various infrared applications, including radiative cooling, thermophotovoltaic devices, thermal camouflage, thermal imaging and biochemical sensing. In this section, we will discuss several examples to illustrate the importance of thermal emitters for infrared applications.

### 9.1 Passive radiative cooling

Passive radiative cooling [171-178] is an emerging technology that can lower temperature below ambient temperature without consuming external energy, offering a promising method to alleviate global warming issue. Spectrally selective thermal emitters, which have high emittance within Earth's atmospheric transparent window (3-5μm and 8-14μm) and simultaneously suppress the infrared absorption outside the window, can passively dissipate heat from Earth into outer space (3 K), thus achieving radiative cooling. Rephaeli et al.

[171] theoretically demonstrated a daytime radiative cooling design through a metal-dielectric photonic structure, as shown in Figure 8(a). Integrated selective emitter and broadband mirror enable enhanced thermal emission within the atmospheric window and reflection of solar light. A net cooling power exceeding 100 W/m$^2$ at ambient temperature is realized. Then, Raman et al. [172] experimentally demonstrated the daytime radiative cooling through a multilayer photonic structure. To meet the large-scale and cost-efficient requirements of commercial applications, Zhai et al. [173] proposed a glass-polymer hybrid metamaterial with silicon dioxide microspheres randomly distributed in transparent polymer to achieve efficient all-day radiative cooling, as shown in Figure 8(b). Benefiting from the phonon enhanced Fröhlich resonances of microspheres, this large-scale design achieves high infrared emissivity within 8 -20 μm and efficient reflection of solar irradiance (∼0.96), enabling an average cooling power of >110 W/m$^2$ continuously for 72 hours. Then, in order to realize more efficient wavelength-selective emission, Li et al. [176] experimentally demonstrated a hierarchically designed polymer nanofibre-based film for all-day radiative cooling, as shown in Figure 8(c). This large-scale design enables selective emission (∼0.78) in the 8-13 μm range and effective sunlight reflection (∼0.96) in the 0.3-2.5 μm range, thus achieving ~3 °C cooling improvement at night and 5 °C sub-ambient cooling under sunlight.

To address the crisis of global warming and energy consumption, personal thermal management (PTM) technologies incorporating passive radiative cooling meta-designs have been extensively investigated. Zeng et al. [177] proposed a hierarchical-morphology metafabric design that contains randomly dispersed scatters, as shown in Figure 8(d). These large-scale woven metafabrics simultaneously exhibit high emissivity (94.5%) in the atmospheric window and high reflectivity (92.4%) in the solar spectrum, enabling a temperature reduction of approximately 4.8°C compared to commercial cotton fabric. By focusing on humans, PTM technologies are promising candidates for efficient energy saving. In contrast to radiative cooling, radiative warming is another energy-saving technology that can be achieved by regulating the infrared emission spectrum to obtain high reflectivity within the atmospheric window and high absorption outside the window [179]. This warming technology provides an effective method to address the global carbon neutrality issue.

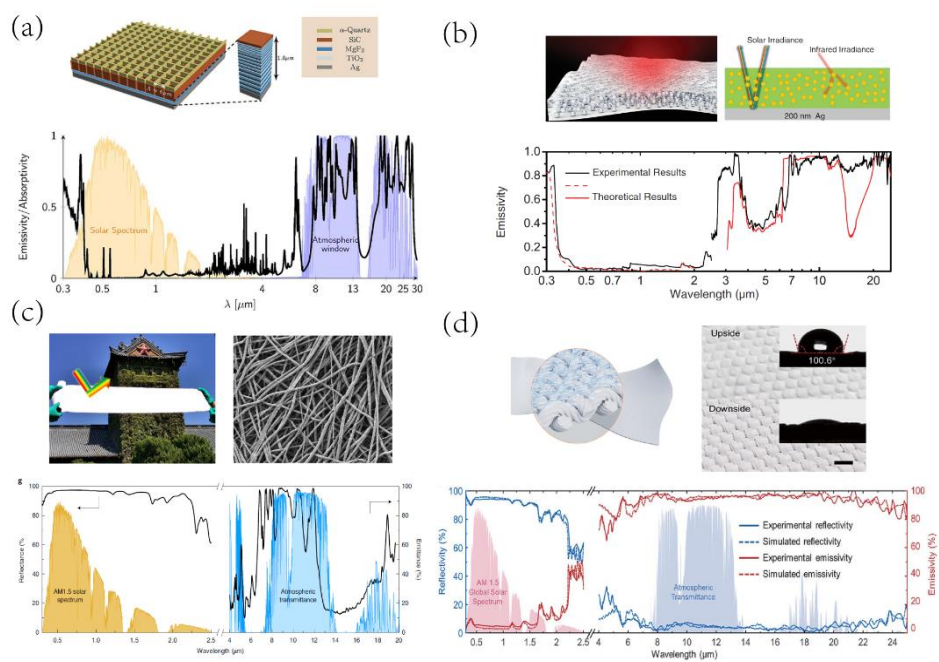

Figure 8. Passive radiative cooling technologies. (a) The top: schematic of the daytime radiative cooler. The bottom: calculated emissivity of the daytime radiative cooler at normal incidence. Reprinted with permission [171]. Copyright 2013, American Chemical Society. (b) The top: the schematic of the polymer-based hybrid radiative cooler with randomly distributed SiO$_2$ microsphere inclusions. The bottom: measured emission spectrum of hybrid radiative cooler compared with theoretical results. Reprinted with permission [173]. Copyright 2017, AAAS. (c) The top: the images of fabricated polymer nanofiber. The bottom: measured emission spectrum of polymer nanofiber film. Reprinted with permission [176]. Copyright 2020, Springer Nature. (d) The top: the schematic of hierarchical-morphology metafabric design. The bottom: measured

emission spectrum of metafabric cooler compared with simulated results. Reprinted with permission [177]. Copyright 2021, AAAS.

## 9.2 Thermophotovoltaic devices

Thermophotovoltaic (TPV) devices [180-191] composed of a heat source, a thermal emitter and a PV cell is a promising technology that converts heat to electricity. The heat source brings the thermal emitter to a high temperature, causing thermal emission, and then the energy is converted to electricity by the PV cell. The heat source can be provided by fuel combustion, nuclear reaction, electrical process and sunlight. Since the photons with energy below the PV cell bandgap cannot be absorbed to generate electricity, a wavelength-selective thermal emitter is necessary to provide emission photons with energy above the PV cell bandgap. As we can see, thermal emitters play an important role in TPV systems to improve the thermal-to-electrical energy conversion efficiency.

Utilizing solar irradiation as a heat source, the energy conversion system that converts sunlight to electricity is known as the solar thermophotovoltaic (STPV) system. A broadband solar absorber is combined in the system to absorb solar photons and get heat up. Chang et al. [190] experimentally demonstrated an STPV device with high thermal stability at temperatures up to at least 1200 °C by combining the refractory W metasurface based solar absorber and thermal emitter, as shown in Figure 9(a). Near-unity absorption in the visible to near-infrared range and suppressed emittance in the long-wave domain are realized to improve the energy conversion efficiency. Overall STPV efficiency as high as 18% is achieved, which is comparable to the efficiency of available commercial single-junction PV cells. In order to further improve the practicality of TPV devices, high-temperature stability and higher TPV efficiency are further required.

## 9.3 Gas and biochemical sensing

Thermal emitters can work as spectrally efficient and low-cost infrared radiation source, which is highly demanded for nondispersive infrared (NDIR) sensing and biochemical analysis. For gas sensing applications [192-196], a thermal emitter can provide a targeted emission peak without the need of an optical filter, thus facilitating the miniaturization of the sensing device. For example, Ref [46] proposed an on-chip gas sensor by combining a microelectromechanical system (MEMS) heater and an infrared thermal emitter. Filter-free, low-cost, and compact gas sensing is realized by utilizing the narrowband thermal emitter as a wavelength-selective infrared radiation source. Furthermore, by integrating metasurface-based emitter and detector within a millimeter-scale nonresonant cavity, a further compact mid-infrared gas sensor is proposed [194]. A $CO_2$ sensitivity of $22.4 \pm 0.5$ ppm·$Hz^{-0.5}$ can be realized, which possesses comparable performance to much larger commercial equipment while consuming 80% less energy per measurement.

On-chip biochemical sensing [197, 198] has also been explored through broadband thermal emitters. The thermal light emitted from the thermal emitter interacting with or transmitting through the detected sample carries vibrational information to the detector. Then through the variations of the measured spectrum with or without the sample, the special vibrational information of the detected sample can be extracted. Compared to conventional spectroscopy devices, an on-chip low-cost thermal source provided by a thermal emitter facilitates the compact infrared spectroscopy applications.

## 9.4 Thermal camouflage

Camouflage is a natural behavior in which creatures such as chameleons, blend into their surroundings in order to hide from predators or prey. Mimicking the natural camouflage, thermal camouflage [199-211] that hides target objects from infrared detectors has attracted increasing interest for many developing infrared devices such as, military and anticounterfeiting equipment. Manipulating surface emissivity is an effective way to prevent thermal emission signal from being detected by various infrared detecting technologies, such as thermal camera. Traditional thermal camouflage can only work at specific background temperatures, making dynamically controllable thermal camouflage more appealing for practical applications. Salihoglu et al. [199] demonstrated a real-time active thermal surface based on multilayer graphene. Through reversible intercalation of nonvolatile ionic liquids, electrically tuning thermal emission can be realized without changing the surface temperature, as shown in Figure 9(b). Moreover, adaptive thermal camouflage is enabled by a feedback mechanism.

With the advancement and diversification of various detection technologies, multispectral thermal camouflage [205] has currently become a hotspot. Camouflage technologies can be combined to confront multiple detectors operating at different wavelength ranges. In the future, by integrating multispectral design with active materials, adaptive thermal camouflage with switchable operating bands may become possible, facilitating the development of adaptive thermal management devices.

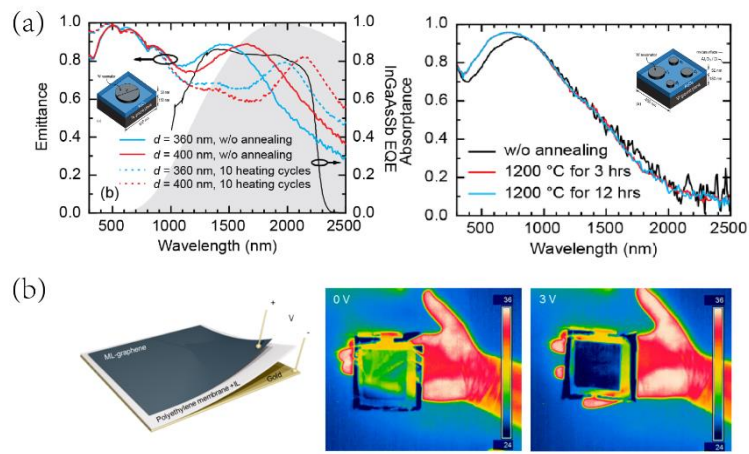

Figure 9. Infrared applications based on thermal emitters. (a) The emission and absorption spectra of W metasurface based thermal emitter (the left) and absorber (the right). Reprinted with permission [191]. Copyright 2018, American Chemical Society. (b) The left: the sketch of multilayer graphene based metasurface design. The right: thermal images of the camouflage design placed on the author's hand under the voltage bias of 0 and 3 V, respectively. Reprinted with permission [200]. Copyright 2018, American Chemical Society.

## 10. Integrated thermal emission microchip based on metasurface array

Thermal emission tuning in multiple degrees of freedom, such as wavelength, bandwidth, polarization, radiation angle and coherence has been extensively investigated. However, these researches are mostly performed on a single metasurface, therefore leading to limited capabilities of tuning thermal emission. Further extending the tuning degrees of freedom is essential for practical infrared applications. Pixelated metasurface array [43, 121, 212] is a promising way to address this obstacle. Integrating multifunctional thermal emitters into a single chip can further facilitate the integration and miniaturization of infrared devices.

Chu et al.[212] experimentally demonstrated a thermal emission microchip with emission wavelengths covering 7-9 μm and 10-14 μm through a pixelated meta-cavity array, as shown in Figure 10(a, c). Each meta-cavity composed of a nanohole metasurface and an FP cavity possesses two emission peaks in the long-wave infrared domain, showing strong x-polarized thermal emission, as shown in Figure 10(b, d). With the increases in the length of the nanoholes, the emission wavelength of meta-cavity gradually redshifts. Through designed nanohole patterns "NJU" and "PHY", polarization, wavelength, and spatial multiplexing thermal emission is further realized and the corresponding thermal imaging results are shown in Figure 10(e). This multiplexing phenomenon comes from the different structure parameters, spatial distribution and nanohole orientations between "NJU" and "PHY". High spatial resolution is enabled in the thermal images with each nanohole performs as an emission element.

Based on this microchip design and the thermal imaging approach, an indirect absorption spectrum measurement technique is further proposed [213], as shown in Figure 10(f). Compared with the traditional absorption spectrum measurement method, this measurement technique is highly compact because it does not require an external infrared radiation source and a complex spectroscopic device. Demonstrated microchip can simultaneously act as a cost-efficient infrared radiation source and the spectroscopic chip. Through the thermal imaging of pixelated meta-cavity array, the spectral absorption information of detected sample can be obtained from spatially distributed meta-cavity pixels. The wavelength resolution of this measurement technology is expected to be improved by utilizing Mie resonances in dielectric metasurfaces or BIC modes [214]. This proposed integrated technology can facilitate on-chip infrared spectroscopy applications, such as gas sensing and microscale biochemical sensors.

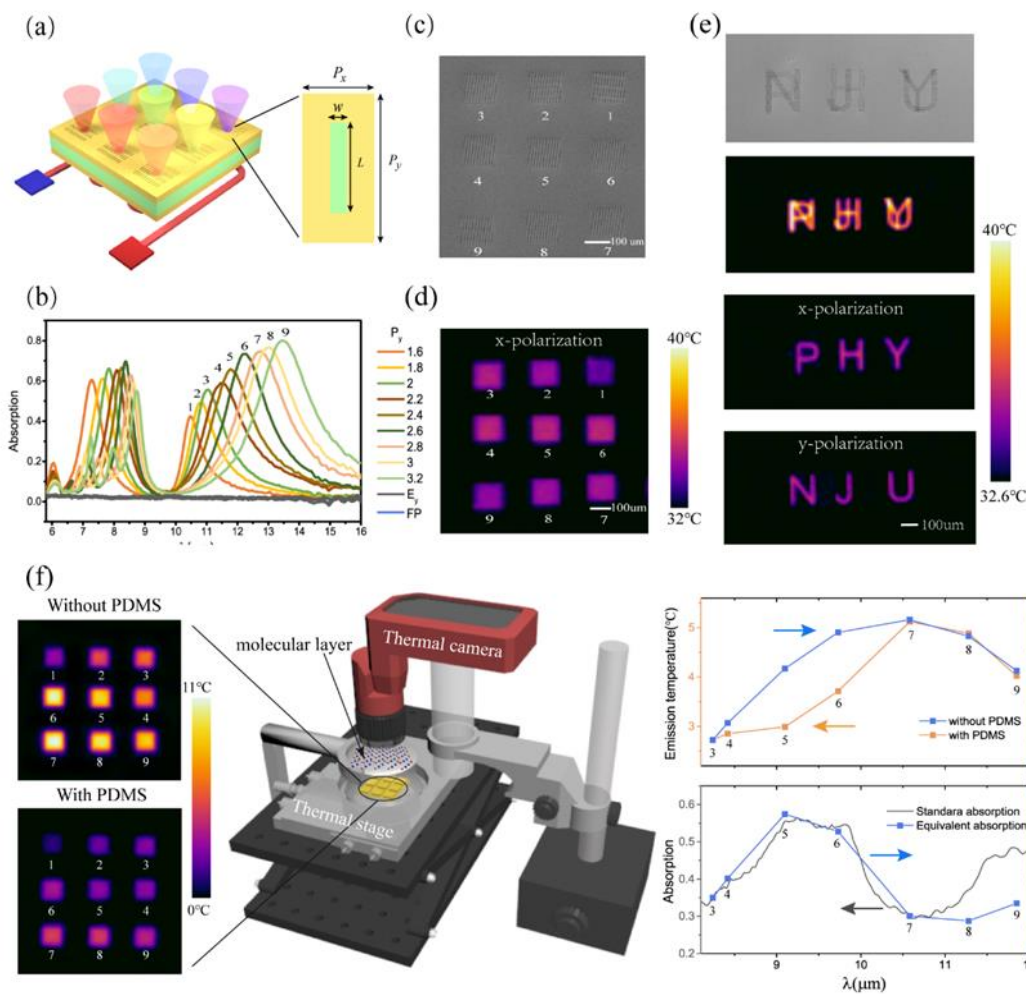

Figure 10. Integrated thermal emission microchip based on a meta-cavity array. (a) The schematic of designed thermal emission microchip composed of 3 × 3 meta-cavity pixels. (b) Measured absorption spectra of meta-cavity array. (c) The SEM picture of fabricated meta-cavity array. (d) The thermal image of the microchip under x polarization. (e) The SEM picture of fabricated nanohole patterns of "NJU" and "PHY". And the thermal images of nanohole patterns. Reprinted with permission [212]. Copyright 2022, De Gruyter. (f) The left: the thermal images of meta-cavity array with and without PDMS layer. The middle: the measurement setup. The right: the emission temperature of meta-cavity pixels and measured equivalent absorption spectrum of PDMS layer. Reprinted with permission [213]. Copyright 2023, © The Optical Society.

## 11. Conclusions

In this review, we have shown that metasurface is a powerful tool for tuning thermal emission in multiple degrees of freedom. Wavelength-selective thermal emission is the key point for improving the efficiency of various thermal management applications. Metasurface-based thermal emitters have successfully achieved the demanded emission spectrum for plenty of infrared devices. Then, the flexible tuning of the radiation angle, polarization, and coherence properties of thermal emission have been respectively discussed. We have also focused on nonreciprocal thermal emission and near-field thermal emission researches, which are important for thermophotovoltaic devices and heat transfer applications. To obtain more tunable degrees of freedom, integrated metasurface array on a single chip can be utilized in a promising way.

Given that integration and miniaturization is the goal for the development of future flat and compact infrared applications, there are still several challenges to be resolved for on-chip thermal emission tuning. 1. Multispectral thermal management designs (from visible to long-wave infrared range) incorporating tuning mechanisms could lead to further integrated devices, facilitating compact multifunctional infrared applications, such as selective stealth in the demanded wavelength range and flexibly switching between multiple functions ranging from anticounterfeiting to radiative cooling. 2. Electrical control is the most common way to obtain high modulation speed of thermal emission. In future, metasurface designs incorporating ultrafast laser technology such as attosecond laser, may bring the modulation speed of tuning thermal emission to the next level, facilitating various dynamic thermal switching devices. 3. High-Q thermal emission is essential for infrared sensors and detecting applications. Further improving the Q factor of the thermal emitter array could bring higher resolution close to the commercial requirement for compact spectroscopy applications. 4. High-temperature stability and the system packaging technology continue to be concerned for energy harvesting and conversion applications.

In recent years, various interesting physics mechanisms have been explored and applied to optical research, such as topological, moiré and non-Hermitian photonics. Combining the topological technologies with the thermal emission tuning mechanisms into metasurface design, various exotic infrared phenomena can be expected. For example, a metasurface possessing BIC modes may provide unique emission properties in momentum space. Moiré photonic designs possessing additional twist angle degrees of freedom, may provide novel effects for twist-angle controlled near-field, far-filed and nonreciprocal thermal emission. The non-Hermitian system may provide EPs, exceptional rings, and even unique Hamiltonian components for controlling thermal emission.

**Author contribution:** All the authors have accepted responsibility for the entire content of this submitted manuscript and approved submission.
**Research funding:**
This work was financially supported by the National Natural Science Foundation of China (Nos. 92163216, 12334015, 92150302, 62288101 and 12004072), the National Key Research and Development Program of China (Grant Nos. 2023YFB2805700), the Natural Science Foundation of Jiangsu Province (BK20200388), and the "Zhishan" Youth Scholar Program of Southeast University (2242021R40013).

**Conflict of interest statement:** The authors declare no conflicts of interest regarding this article.